\documentclass[aps,prc,twocolumn,superscriptaddress,showpacs]{revtex4}
\usepackage{amsmath,graphicx}
\usepackage{dcolumn} 

\newcommand{\mean}[1]{\left\langle #1 \right\rangle} 

\begin{document}

\preprint{ULB-TH/02-03}

\title{Effects of momentum conservation on the analysis of anisotropic
flow}

\author{N. Borghini}
\email{Nicolas.Borghini@ulb.ac.be}
\affiliation{Service de Physique Th{\'e}orique, CP 225, 
Universit{\'e} Libre de Bruxelles, B-1050 Brussels, Belgium}

\author{P. M. Dinh}
\email{dinh@spht.saclay.cea.fr}

\author{J.-Y. Ollitrault}
\email{Ollitrault@cea.fr}
\altaffiliation[also at ]{L.P.N.H.E., Universit{\'e} Pierre et Marie Curie, 
4 place Jussieu, F-75252 Paris cedex 05, France}

\affiliation{Service de Physique Th{\'e}orique, CEA-Saclay, 
F-91191 Gif-sur-Yvette cedex, France}

\author{A. M. Poskanzer}
\email{AMPoskanzer@LBL.gov}
\affiliation{Nuclear Science Division, Lawrence Berkeley National
Laboratory, Berkeley CA 94720}

\author{S. A. Voloshin}
\email{Voloshin@physics.wayne.edu}
\affiliation{Department of Physics and Astronomy, Wayne State 
University, 666 W. Hancock Street, Detroit, MI 48202}

\date{\today}

\begin{abstract}
We present a general method for taking into account correlations due
to momentum conservation in the analysis of anisotropic flow, either by 
using the two-particle correlation method or the standard flow vector method. 
In the latter, the correlation between the particle and the flow vector is 
either corrected through a redefinition (shift) of the flow vector, or 
subtracted explicitly from the observed flow coefficient. 
In addition, momentum conservation contributes to the reaction plane 
resolution.
Momentum conservation mostly affects the first harmonic in 
azimuthal distributions, i.e., directed flow.  It also modifies higher
harmonics, for instance elliptic flow, when they are measured with
respect to a first harmonic event plane such as one determined with
the standard transverse momentum method.  
Our method is illustrated by application to NA49 data on pion directed flow.
\end{abstract}


\maketitle

\section{Introduction}

In heavy ion collisions, the transverse momenta of outgoing particles are 
correlated with the direction of the impact parameter  (reaction plane) of the
two incoming nuclei.  For example, at energies above 100~MeV per nucleon,
nucleons in the projectile rapidity region are deflected
away from the target.  This collective effect is usually referred to
as directed (or sidewards) flow.  
It is characterized by the
first Fourier coefficient of the particle azimuthal distribution,
$v_1$~\cite{Voloshin:1996mz}:
\begin{equation}
v_1\equiv\mean{\cos(\phi - \Phi_{RP})},
\end{equation}
where $\phi$ denotes the azimuthal angle of an outgoing particle,
$\Phi_{RP}$ is the azimuth of the reaction plane, and angular brackets
denote an average over events and over all particles in 
a given transverse momentum and rapidity window. 
{\em Differential} flow is characterized by $v_1(p_T,y)$ 
that describes the directed flow in a narrow
transverse momentum ($p_T$) and rapidity ($y$) interval.

The most common observable to quantify directed flow, introduced by
Danielewicz and Odyniec in 1985~\cite{Danielewicz:1985hn}, is the mean
transverse momentum projected on the reaction plane, $\mean{p_x}$, as
a function of rapidity.  In terms of $v_1$, it can be written as
$\mean{p_x}=\mean{p_T\,v_1(p_T,y)}$, where the average in the
right-hand side is taken over $p_T$.  This observable is believed to
be sensitive to the nuclear compressibility and to in-medium cross
sections~\cite{Magestro:ba}.  Experimentally, $\mean{p_x}$ is
extracted from azimuthal correlations between the outgoing particles,
under the assumption that these correlations are only due to flow.
However, it was soon realized that global momentum conservation also
induces azimuthal correlations.  
Different ways to correct for this effect were devised in 
Refs.~\cite{Danielewicz:1988in,ogilvie}.  These methods are still 
in use at intermediate~\cite{Cussol:2001df,Prendergast:2000fv} as well as 
at relativistic energies~\cite{Ahle:jv,E895,Crochet:1997hz}. 
Generally, corrections are important for weak flow 
and/or small systems, when there are physical or detector differences
between the forward and backward hemispheres in a reaction.

More detailed information can be obtained on the physics involved in
the collision by studying the transverse momentum dependence of
directed flow~\cite{Pan:1992ef}, $v_1(p_t)$ in a given rapidity window.  
Detailed theoretical studies
have been presented for this differential directed flow
\cite{Li:1996iz,Li:1999bh,Li:2000bj}.  This was accompanied by new
analyses of experimental data, and a wealth of results are now
available from FOPI at SIS~\cite{FOPI}, E877 at AGS~\cite{E877} and
NA49 at CERN~\cite{NA49}.  In these analyses, unfortunately,
correlations arising from momentum conservation are not taken into
account.  Recently, it was shown~\cite{Borghini:2000cm} that they may
in fact be of the same order of magnitude as correlations due to flow,
leading to a reevaluation of the ``standard'' flow analysis results.

In this paper, we show how to modify standard flow analysis techniques
in order to take into account momentum conservation.  Two flow methods
are used experimentally.  The first one relies on a study of
two-particle correlations~\cite{Wang:1991qh}.  This simple case is
discussed in Sec.~\ref{s:twoparticle}.  The second method, which is by
far the most common, involves the construction of a ``flow vector''
used as an estimate of the reaction
plane~\cite{Danielewicz:1985hn,Poskanzer:1998yz}.  In order to measure
the flow of a particle, one correlates it to the estimated event
plane of the flow vector, and corrects
this correlation by dividing by a ``resolution'' factor which
takes into account the uncertainty of the reaction plane
determination.  
This procedure is recalled in Sec.~\ref{s:standard}.  
In Sec.~\ref{s:correction}, we show how this method can be modified to
take into account momentum conservation.  

The momentum conservation leads to two effects, whose magnitude
is controlled by a parameter $f$ which is roughly the
square root of the fraction of all particles used in the estimate
of the event plane.  First, momentum conservation
contributes to the correlation between a particle and the flow vector,
proportionally to the parameter $f$ (Sec.~\ref{s:subtraction}).  
This spurious correlation can be eliminated through a redefinition of 
the flow vector (Sec.~\ref{s:shift}). 
In addition, momentum conservation affects the
resolution of the event plane, although the effect is smaller, as it
is quadratic in $f$ (Sec.~\ref{s:resolution}).  
This second effect may also bias higher
harmonics measured with respect to the first harmonic event plane.
(Note, however, that momentum conservation has 
no effect when elliptic flow is measured with
respect to the second harmonic event plane.) Our results are discussed
in Sec.~\ref{s:discussion}; in particular, we show that the
correlation from momentum conservation vanishes if the detector
acceptance is symmetric with respect to midrapidity, because particles
belonging to the forward and backward hemispheres are given weights
with opposite signs.  In Sec.~\ref{s:experiment}, the method is
illustrated by NA49 data on pion directed flow.  Finally, technical
details are exposed in the Appendixes \ref{s:appA} and \ref{s:appB}.

\section{Two-particle correlation technique}
\label{s:twoparticle}

Since the orientation of the reaction plane is not known, $v_1$ is
extracted from azimuthal correlations between outgoing particles.  The
simplest way consists in using two-particle azimuthal
correlations~\cite{Wang:1991qh}.  These are related to $v_1$ by the
identity:
\begin{eqnarray}
\label{twoflow}
\mean{\cos(\phi_1 - \phi_2)} & = & 
\mean{\cos(\phi_1 - \Phi_{RP})\cos(\phi_2 - \Phi_{RP})}\cr
&=&\mean{\cos(\phi_1 - \Phi_{RP})}\mean{\cos(\phi_2 - \Phi_{RP})}\cr
&=& v_1(1)\,v_1(2),
\end{eqnarray}
where $v_1(1)$ and $v_1(2)$ denote the flow coefficients associated
with each particle.  For spherical nuclei, symmetry with respect to
the reaction plane ensures that the $\mean{\sin(\phi- \Phi_{RP})}$
terms are zero. In going from the first to the second line, we have
assumed that all azimuthal correlations are due to flow or,
equivalently, that azimuthal angles with respect to the reaction plane
$\phi_1- \Phi_{RP}$ and $\phi_2- \Phi_{RP}$ are statistically independent.  
Recent implementations of the two-particle correlation method can be found in
Refs.~\cite{Prendergast:2000fv,Lacey:2001}.

As recalled in the introduction, however, there are also azimuthal
correlations induced by global momentum conservation, so that the
measured correlation is in fact the sum of two terms(when both effects are 
small)
\begin{equation}
\label{twoflow+mom}
\mean{\cos(\phi_1-\phi_2)}= 
v_1(1) v_1(2)+ \mean{\cos(\phi_1-\phi_2)}_{\rm mom.\ cons.},
\end{equation}
where the last term, due to momentum conservation, is evaluated in
Appendix A of Ref.~\cite{Borghini:2000cm}:
\begin{equation}
\label{twomomentum}
\mean{\cos(\phi_1-\phi_2)}_{\rm mom.\ cons.}= 
-\frac{{p_T}_1{p_T}_2}{N\mean{p_T^2}},
\end{equation}
where $N$ denotes the {\em total} number of particles produced in the
collision, and the average value $\mean{p_T^2}$ is similarly taken
over all produced particles. 
Since one usually detects only a fraction of
the produced particles, the denominator of Eq.~(\ref{twomomentum}) should
be estimated using a model, or by an extrapolation of available data
to full phase space.

In order to take into account the correlations due to momentum
conservation in the analysis of directed flow, one simply subtracts
the contribution due to momentum conservation, Eq.\ (\ref{twomomentum}), 
 from the measured
two-particle correlation Eq.\ (\ref{twoflow+mom}), so as to isolate
the correlation due to flow.  This is the procedure followed in
Ref.~\cite{Borghini:2000cm} to estimate the effects of momentum
conservation on the analysis of directed flow at SPS.

\section{Flow-vector method}
\label{s:standard}

Most analyses of directed flow rely on the construction of the event 
flow-vector~\cite{Danielewicz:1985hn,Poskanzer:1998yz} defined as:
\begin{equation}
\label{Qvector}
{\bf Q} = \sum_{j=1}^N w_j {\bf u}_j,
\end{equation}
where the sum runs over all produced particles, ${\bf u}_j$ denotes
the unit vector of the particle transverse momentum, ${\bf u}_j={\bf
e_x}\cos\phi_j+{\bf e_y}\sin\phi_j={\bf p_T}_j/{p_T}_j$, and $w_j$ is
a weight depending on the particle type, its transverse momentum
${p_T}_j$ and rapidity $y_j$.  
The weights, $w_j$, are usually given non zero values only in regions of good 
detector acceptance.

The azimuthal angle of the flow vector, $\Phi$, is considered an
estimate of the reaction plane azimuth, $\Phi_{RP}$.  Therefore, the
azimuthal angle of ${\bf Q}$ with respect to the reaction plane is
usually referred to as the uncertainty in the reaction plane determination,
and will be denoted by $\Delta\Phi\equiv \Phi-\Phi_{RP}$.  
One then writes an equation
similar to Eq.~(\ref{twoflow}) for the azimuthal correlation between a
particle with azimuth $\phi$ and the flow vector:
\begin{eqnarray}
\label{flowQ}
\mean{\cos(\phi-\Phi)} &=&
\mean{\cos(\phi-\Phi_{RP}-\Phi+\Phi_{RP})} \cr
&=& \mean{\cos(\phi-\Phi_{RP})}\mean{\cos\Delta\Phi} \cr 
&=& v_1 \mean{\cos\Delta\Phi}. 
\end{eqnarray}
This equation once again relies on the assumption that azimuthal
correlations are only due to flow.
To avoid autocorrelation one should subtract the contribution of the 
particle under study from the flow vector.
Since the flow vector,
Eq.~(\ref{Qvector}), involves a summation over many particles, the
correlation between a particle and the flow vector, Eq.\
(\ref{flowQ}), is usually much stronger than the correlation between
two particles, Eq.\ (\ref{twoflow}).  
This is one of the main reasons why the flow vector
method is more commonly used.  However, correlations due to momentum
conservation are also stronger, so that corrections due to momentum
conservation are of the same magnitude with both methods, as we shall
see in Sec.~\ref{s:discussion}. In the rest of this paper $\Phi_{RP}$
will be taken as zero; that is, the reaction plane is along the
$x$-axis.

The next step in the analysis is to evaluate the ``resolution''
$\mean{\cos\Delta\Phi}$.  This factor results from a competition
between flow, which tends to align ${\bf Q}$ in the direction of the
reaction plane, and statistical fluctuations, whose relative magnitude
decreases like $1/\sqrt{N}$.  A quantitative estimate can be obtained
with help of the central limit theorem.  This theorem allows one to
write the distribution of the flow vector as a Gaussian distribution
in the coordinate system where the orientation of the reaction plane
is fixed~\cite{Voloshin:1996mz,Ollitrault:1992bk}:
\begin{equation}
\label{distri-Q}
\frac{dP}{d{\bf Q}}=\frac{1}{\pi\sigma^2}
\exp\left(-\frac{({\bf Q}-\mean{\bf Q})^2}{\sigma^2}\right).
\end{equation}
The average value $\mean{\bf Q}$ is aligned with the reaction plane,
i.e., with the $x$-axis, while the dispersion $\sigma$ is due to
statistical fluctuations.  Writing ${\bf Q}=Q({\bf
e_x}\cos\Delta\Phi+{\bf e_y}\sin\Delta\Phi)$, and integrating over $Q$
and $\Delta\Phi$, one
obtains~\cite{Poskanzer:1998yz,Ollitrault:1997di}:
\begin{equation}
\label{resolution}
\mean{\cos\Delta\Phi}= \frac{\sqrt{\pi}}{2} \chi\,e^{-\chi^2/2}
\left[I_0(\chi^2/2)+I_1(\chi^2/2)\right],
\end{equation}
where $I_0$, $I_1$ are modified Bessel functions and $\chi$ is the
dimensionless parameter $\chi\equiv \mean{\bf Q}\cdot{\bf
e_x}/\sigma$, which characterizes the relative magnitude of flow and
statistical fluctuations. In what follows, $\chi$ will be referred to
as the ``resolution parameter'' \footnote{Note that we follow the
definition introduced in Refs.~\cite{Ollitrault:1997di,Ollitrault:ba},
while the value of $\chi$ adopted in
Ref.~\cite{Voloshin:1996mz,Poskanzer:1998yz} is larger by a factor of
$\sqrt{2}$.}.

This parameter $\chi$ is usually estimated from the correlation
between two ``subevents'' \cite{Danielewicz:1985hn}:
the set of $N$ particles is divided randomly into two equivalent
subsets of $N/2$ particles, labeled $a$ and $b$.  One then constructs
two flow vectors ${\bf Q_a}$ and ${\bf Q_b}$ defined as in
Eq.~(\ref{Qvector}), with the summation running over the corresponding
subset.  Flow induces an azimuthal correlation between ${\bf Q_a}$ and
${\bf Q_b}$, since both are correlated with the reaction plane.  
Then, when all azimuthal
correlations are due to flow, the distribution of $\Phi_a-\Phi_b$
depends only on the resolution parameter $\chi$.  
To find the functional form of this relationship, 
one writes the probability distribution of
${\bf Q_a}$ in a form similar to Eq.\ (\ref{distri-Q}):
\begin{equation}
\label{distsubeva}
\frac{dP}{d{\bf Q_a}} = \frac{1}{\pi\sigma_a^2} 
\exp \left(-\frac{({\bf Q_a}-\mean{\bf Q_a})^2}{\sigma_a^2} \right), 
\end{equation}
and a similar formula for ${\bf Q_b}$ with 
$\mean{\bf Q_b}=\mean{\bf Q_a}\equiv \mean{\bf Q_{\rm sub}}$ and 
$\sigma_b=\sigma_a\equiv \sigma_{\rm sub}$ since the subevents are equivalent.  
The distribution of the relative angle $dP/d(\Phi_a-\Phi_b)$ depends only
on the resolution parameter associated with either subevent
$\chi_{\rm sub}\equiv\mean{\bf Q_{\rm sub}}\cdot{\bf e_x}/\sigma_{\rm sub}$.  Its analytic
expression can be found in
Refs.~\cite{Ollitrault:1997di,Ollitrault:ba}, and provides a perfect
fit to experimental data~\cite{Taranenko:1999yh}.  The parameter
$\chi_{\rm sub}$ is usually obtained from the square root of the average
cosine of the relative angle, $\sqrt{\mean{\cos(\Phi_a-\Phi_b)}}$,
which is simply the resolution of a subevent plane, and is given by
Eq.~(\ref{resolution}) with $\chi$ replaced by $\chi_{\rm sub}$.  Equation
(\ref{resolution}) must then be inverted in order
to obtain $\chi_{\rm sub}$~\cite{Poskanzer:1998yz}.  
In a simpler way, $\chi_{\rm sub}$ can be obtained directly from the 
fraction of events with the relative angle between subevents 
$|\Phi_a-\Phi_b|$ larger than $\pi/2$~\cite{Ollitrault:1997di,Ollitrault:ba}:
\begin{equation}
\label{obtus}
P(|\Phi_a-\Phi_b|>\pi/2)=\frac{1}{2} e^{-\chi_{\rm sub}^2}. 
\end{equation}
This method is used for example in Ref.~\cite{Taranenko:1999yh}. If
$\chi_{\rm sub}$ is significantly larger than unity, however, only a small
fraction of events have $|\Phi_a-\Phi_b|$ larger than $\pi/2$, and the
resulting statistical error on $\chi_{\rm sub}$ is larger than with the
previous method.  In the limiting case $\chi_{\rm sub}\gg 1$, the distribution
of the relative angle $\Phi_a-\Phi_b$ is a Gaussian, and $\chi_{\rm sub}$ can
be directly obtained from its width,
$\mean{(\Phi_a-\Phi_b)^2}=1/\chi_{\rm sub}^2$.

The last step is to relate the parameter $\chi_{\rm sub}$, corresponding to
one subevent, to $\chi$, corresponding to the whole event.  Since each
subevent contains half of the particles, $\mean{\bf Q_{\rm sub}}=\mean{\bf
Q}/2$.  If correlations between particles are only due to flow, a
similar scaling occurs in the statistical fluctuations:
$\sigma_{\rm sub}^2=\sigma^2/2$, so that
\begin{equation}
\label{chivschia}
\chi=\chi_{\rm sub}\sqrt{2}. 
\end{equation}
The flow-vector method is defined by Eqs.~(\ref{flowQ}),
(\ref{resolution}), (\ref{chivschia}), and a technique for obtaining
$\chi_{\rm sub}$.  Equations~(\ref{flowQ}), (\ref{obtus}), and
(\ref{chivschia}), must be modified when taking into account momentum
conservation.  This is studied in Sec.~\ref{s:correction}.

The first harmonic flow vector, Eq.~(\ref{Qvector}), can also be used
to measure higher harmonics of the azimuthal distribution
$v_n=\mean{\cos n\phi}$ for $n>1$, with the help of an equation
similar to Eq.~(\ref{flowQ}):
\begin{equation}
\label{flownQ}
\mean{\cos n(\phi-\Phi)}= v_n \mean{\cos n\Delta\Phi}. 
\end{equation}
The corresponding resolution $\mean{\cos n\Delta\Phi}$ depends on the
same parameter $\chi$~\cite{Poskanzer:1998yz,Ollitrault:1997di}.  This
method has been used to analyze elliptic flow $v_2$ \cite{elliptic}
relative to a first harmonic event plane.  Since, as we have mentioned
above, momentum conservation modifies the value of $\chi$, it must
also be taken into account in this case.  We come back to this
discussion in Sec.~\ref{s:discussion}.

\section{Modifications due to momentum conservation}
\label{s:correction}

Global momentum conservation induces correlations between outgoing
particles and the flow vector.  In the same way as Eq.~(\ref{twoflow})
is replaced by Eq.~(\ref{twoflow+mom}), Eq.~(\ref{flowQ}) becomes
\begin{equation}
\label{flowQ+mom}
\mean{\cos(\phi-\Phi)}= v_1 \mean{\cos\Delta\Phi} + 
\mean{\cos(\phi-\Phi)}_{\rm mom.\ cons.}.
\end{equation}
There are two methods to take into account the additive term. 
The first one consists in redefining the ${\bf Q}$-vector [i.e., modify $\Phi$ 
in Eq.~(\ref{flowQ+mom})] in each event, in such a way that 
the additional term no longer appears (Sec.~\ref{s:shift}).
Alternatively, one can compute explicitly the additive term
and subtract it in order to isolate the correlation due to flow
(Sec.~\ref{s:subtraction}). 
Furthermore, momentum conservation affects the resolution 
$\mean{\cos\Delta\Phi}$, and the methods used to estimate the resolution 
must be modified accordingly (Sec.~\ref{s:resolution}).

\subsection{Shifted flow-vector method}
\label{s:shift}

The correlation due to momentum conservation between the flow vector 
and a particle with momentum ${\bf p_T}$ can be characterized 
by the average value of ${\bf u}\cdot{\bf Q}$, where 
${\bf u}\equiv {\bf p_T}/p_T$.  
Using the definition of the ${\bf Q}$-vector, Eq.~(\ref{Qvector}), and 
the expression of the two-particle correlation due to momentum conservation, 
Eq.~(\ref{twomomentum}), one obtains:
\begin{equation}
\label{avdotproduct}
\mean{{\bf u}\cdot{\bf Q}}_{\rm mom.cons.}=
- p_T\,\frac{\mean{w p_T}}{\mean{p_T^2}},
\end{equation}
where averages in the right-hand side run over {\em all} the produced
particles.  
This suggests one could redefine the flow vector as 
\begin{equation}
\label{shiftedQ}
{\bf Q'}={\bf Q}+ \frac{\mean{w p_T}}{\mean{p_T^2}} {\bf p_T},
\end{equation}
so that momentum conservation does not contribute to the correlation 
between ${\bf Q'}$ and the particle under study. 
As a matter of fact, when there is no flow the shifted flow vector ${\bf Q'}$ 
and ${\bf p_T}$ are now uncorrelated by momentum conservation: 
any correlation between ${\bf Q'}$ and ${\bf p_T}$ arises from flow, because 
here and throughout the paper, we neglect other sources of nonflow 
correlations. 

A similar method was proposed in Ref.\ \cite{ogilvie}, in the context of 
intermediate-energy collisions, where the shifted flow vector was defined as
\begin{equation}
\label{Q'Ogilvie}
{\bf Q'}={\bf Q}+ \frac{\sum_j w_j m_j/{p_T}_j}{\sum_j m_j} {\bf p_T},  
\end{equation}
where $m_j$ is the mass of particle $j$. 
If $\mean{p_T^2}$ of different particles scales with their mass $m$, 
our definition Eq.\ (\ref{shiftedQ}) 
reduces approximately to Eq.\ (\ref{Q'Ogilvie}). 
That is the case in particular for a thermalized nonrelativistic gas at rest 
with (almost) no flow, as in Ref.\ \cite{ogilvie}. 
However, Eq.\ (\ref{Q'Ogilvie}) is no longer valid when there is flow, 
and/or relativistic effects, whereas Eq.\ (\ref{shiftedQ}) still holds.

\subsection{Explicit subtraction}
\label{s:subtraction}

Instead of modifying the flow vector event-by-event, 
one can as well keep the original definition, Eq.\ (\ref{Qvector}), 
and subtract the contribution of momentum conservation to the 
measured correlations, Eq.\ (\ref{flowQ+mom}), at the end of 
the analysis. 

Let us first estimate the order of magnitude of the last term of 
Eq.\ (\ref{flowQ+mom}).  We
assume for simplicity that the particle under study has no flow,
$v_1=0$, so that only the correlation due to momentum conservation
remains.  We then make the following approximation:
\begin{equation}
\label{approximation}
\mean{\cos (\phi-\Phi)}_{\rm mom.\ cons.}= 
\mean{\frac{{\bf u}\cdot{\bf Q}}{Q}}
\sim \frac{\mean{{\bf u}\cdot{\bf Q}}}{\mean{Q}},
\end{equation}
with ${\bf u}\equiv {\bf p_T}/p_T$.  
The average dot product is given by Eq.(\ref{avdotproduct}). 
Furthermore, the average length of the flow vector
$\mean{Q}$ is approximately equal to the r.m.s.\ length
$\sqrt{\mean{Q^2}}$.  Neglecting correlations between momenta, one
obtains from the definition, Eq.\ (\ref{Qvector}), $\mean{Q^2}\sim
N\mean{w^2}$.  All in all, the order of magnitude of the correlation
due to momentum conservation, Eq.~(\ref{approximation}), reads
\begin{equation}
\label{ordermag}
\mean{\cos(\phi-\Phi)}_{\rm mom.\ cons.}\sim 
-\frac{p_T}{\sqrt{N\mean{p_T^2}}}\,f,
\end{equation}
where $f$ is a dimensionless quantity given by 
\begin{eqnarray}
\label{deff}
f &\equiv& \frac{\mean{wp_T}}{\sqrt{\mean{w^2}\mean{p_T^2}}}\cr
&=& \mean{wp_T}_Q \sqrt{\frac{M}{\mean{w^2}_Q N \mean{p_T^2}}},
\end{eqnarray}
and the subscript $Q$ refers to those $M$ particles used for the ${\bf
Q}$-vector
(for which weight $w_i \ne 0$). 
The weight $w$ vanishes for particles not
seen in the detector, so that $\mean{w^2}$ and $\mean{wp_T}$ scale
like the fraction of {\em detected} particles. Hence, $f$ scales like
the square root of the fraction of detected particles.  This is seen
more clearly in the second line of Eq.~(\ref{deff}) where there is a
factor $\sqrt{M/N}$. If $w$ is taken as $+1$ for particles detected in
the forward hemisphere and $-1$ for the backward hemisphere, then
$\mean{wp_T}_{Q} = (M_F \mean{p_T}_F - M_B \mean{p_T}_B)/M$, where $F$
and $B$ refer to particles used for the $\bf{Q}$-vector from the
forward and backward hemispheres, respectively, so that $M = M_F +
M_B$. In this case the quantity $\mean{w^2}_Q = 1$.

The estimate in Eq.~(\ref{ordermag}) is only an order of magnitude.  A
more accurate expression can be obtained under very general
assumptions.  We show in Appendix \ref{s:appA} that the actual value
of the correlation due to momentum conservation between a particle and the 
flow vector is
\begin{eqnarray}
\label{correl}
\lefteqn{\mean{\cos(\phi-\Phi)}_{\rm mom.\ cons.} =} & \cr
 &\displaystyle -\frac{p_T}{\sqrt{N\mean{p_T^2}}}\,\frac{f}{\sqrt{1-f^2}}\, 
\frac{\sqrt{\pi}}{2}e^{-\chi^2/2}I_0(\chi^2/2), 
\end{eqnarray}
where $\chi\equiv \mean{\bf Q}\cdot{\bf e_x}/\sigma$ has been defined
in Sec.~\ref{s:standard}.  It is worth noting that the Eq.\
(\ref{correl}) result now depends on $\chi$, expressing the effect of
flow in reducing the correlation due to momentum
conservation. However, the dependence is weak for small $\chi$ (i.e.,
in the case of poor resolution).

\subsection{Correction to the reaction plane resolution}
\label{s:resolution}

We now discuss the influence of momentum conservation on the
determination of the resolution parameter $\chi$.  
The corrections discussed below hold for the method of 
Sec.~\ref{s:shift} as well as for the method of Sec.~\ref{s:subtraction}.

As recalled in
Sec.~\ref{s:standard}, the resolution is determined from the
correlation between subevents, i.e., from the distribution of the
relative angle $\Phi_a-\Phi_b$ between the subevents.  We show in
Appendix \ref{s:appB} that momentum conservation contributes to this
correlation such that the distribution of $\Phi_a-\Phi_b$ now depends
both on the resolution parameter of the subevent $\chi_{\rm sub}$, and on the
parameter $f$ introduced in Eq.~(\ref{deff}), in a non trivial way:
taking into account momentum conservation does not only amount to
changing the value of $\chi_{\rm sub}$.  This is because the shape of the
distribution of the relative angle $\Phi_a-\Phi_b$ depends on whether
the correlation is due to flow or to other effects.  This is
illustrated in Ref.~\cite{Ollitrault:dy}, where the distribution of
$\Phi_a-\Phi_b$ is plotted in the two limiting cases when the
correlations are only due to flow ($f=0$), or only to other, nonflow
effects ($\chi_{\rm sub}=0$).

As recalled in Sec.~\ref{s:standard}, there are several methods of
extracting $\chi_{\rm sub}$ from the distribution of $\Phi_a-\Phi_b$.  The
usual method relies on the calculation of
$\langle\cos(\Phi_a-\Phi_b)\rangle$.  If $f$ is much smaller than
unity, Eq.~(\ref{resolution}) becomes:
\begin{eqnarray}
\label{resolution2}
\lefteqn{\mean{\cos(\Phi_a-\Phi_b)}^{1/2} =} & \cr
& \displaystyle \frac{\sqrt{\pi}}{2} \,e^{-\chi_{\rm sub}^2/2} \left[
\chi_{\rm sub}^2\left[I_0(\chi_{\rm sub}^2/2)+I_1(\chi_{\rm sub}^2/2)\right]^2\right. \cr
& \displaystyle -\left.
\frac{f^2}{2}\left[I_0^2(\chi_{\rm sub}^2/2)+I_1^2(\chi_{\rm sub}^2/2)\right]
\right]^{1/2},
\end{eqnarray}
Once the left-hand side and $f$ are known, $\chi_{\rm sub}$ can be obtained by
solving this equation.  
This expression holds only
to leading order in $f$. We were not able to obtain an analytic
formula for arbitrary $f$.

It may be 
simpler to extract $\chi_{\rm sub}$ from the fraction of events
for which $|\Phi_a-\Phi_b|>\pi/2$. Then, the correction for momentum
conservation can be implemented using the following exact formula,
derived in Appendix~\ref{s:appB}:
\begin{equation}
\label{obtuscorr}
P(|\Phi_a-\Phi_b|>\pi/2)=\frac{1}{2-f^2} e^{-\chi_{\rm sub}^2}, 
\end{equation}
instead of Eq.~(\ref{obtus}).  If the resolution parameter is large,
$\chi_{\rm sub} > 1$, this method leads to larger statistical errors as
recalled in Sec.~\ref{s:standard}.  If $\chi_{\rm sub}\gg 1$, the
distribution of $\Phi_a-\Phi_b$ is Gaussian and $\chi_{\rm sub}$ can be
obtained from its width:
\begin{equation}
\label{highresolution}
\mean{(\Phi_a-\Phi_b)^2}=\frac{2}{\chi_{\rm sub}^2 (2-f^2)}.
\end{equation}

Finally, the relation between the resolution parameter of the
subevent, $\chi_{\rm sub}$, and that of the whole event, $\chi$, is easily
obtained using Eqs.~(\ref{defsigma}), (\ref{defsigmaa}), and the
scaling relation $\mean{\bf Q_{\rm sub}}=\mean{\bf Q}/2$:
\begin{equation}
\label{chi&chia}
\chi = \chi_{\rm sub}\sqrt{\frac{2-f^2}{1-f^2}}.
\end{equation}

When correlations are weak (i.e., when both $f$ and $\chi$ are small compared 
to unity), the event plane resolution can be corrected for momentum conservation 
through the following simple formula:
\begin{equation}
\label{weakcorr}
\chi^2({\rm corrected})=\chi^2({\rm uncorrected})+f^2.
\end{equation}
This approximation can be used instead of 
Eqs.\ (\ref{resolution2})/(\ref{obtuscorr}) and (\ref{chi&chia}) on the $\chi$ 
of the full event plane as obtained in Sec. \ref{s:standard}.
%

Let us summarize the results obtained in this Section.  The crucial
parameter, which determines the magnitude of corrections due to
momentum conservation, is the dimensionless parameter 
$f$, defined in
Eq.~(\ref{deff}). 
Once this quantity has been estimated, 
the subevent plane resolution parameter, $\chi_{\rm sub}$, 
can be obtained from either
$\mean{\cos(\Phi_a-\Phi_b)}$ using Eq.~(\ref{resolution2}), if $f\ll
1$, or 
from the fraction of subevent angle
differences greater than $\pi/2$ using Eq.~(\ref{obtuscorr}).  Or, if
the resolution parameter is very large, from
Eq.~(\ref{highresolution}).  
The full event plane
resolution parameter, $\chi$, is obtained from Eq.~(\ref{chi&chia}),
and  
the full event plane resolution from Eq.~(\ref{resolution}).
Finally, 
subtracting from the measured values of $\mean{\cos(\phi - \Phi)}$
the contribution due to momentum conservation, Eq.~(\ref{correl}),
and dividing the result by the 
event plane 
resolution, 
one obtains 
the final $v_1$ values. 
When $f$ vanishes, one recovers the formulas of the standard
flow analysis, recalled in Sec.~\ref{s:standard}.

\section{Discussion}
\label{s:discussion}

Corrections arising from momentum conservation are twofold.  First,
they contribute to the correlation between a particle and the event
plane. This contribution, which increases linearly with the particle
transverse momentum~\cite{Borghini:2000cm}, must be subtracted from
the measured correlation so as to isolate the correlation due to flow;
see Eq.~(\ref{flowQ+mom}).  Next, momentum conservation affects the
resolution of the event plane, which results in a global
multiplicative factor.  Both effects must be taken into account in the
analysis of directed flow, but only the second one in the case of
higher harmonics relative to a first harmonic event plane.

All effects of momentum conservation disappear if the parameter $f$
defined in Eq.~(\ref{deff}) vanishes, which occurs if $\mean{wp_T}=0$.
Since $w$ is usually an odd function of the center-of-mass
rapidity~\cite{Danielewicz:1985hn}, $\mean{wp_T}$ vanishes as soon as
the detector acceptance is symmetric with respect to midrapidity, as
was already mentioned in Refs.~\cite{Danielewicz:1988in,Poskanzer:1998yz}.
Another 
theoretically 
interesting limit is the case when all particles are detected
and, in particular, when $w=p_T$, which implies $f=1$. In this case,
the flow vector, Eq.~(\ref{Qvector}), is the sum of all transverse
momenta, which vanishes due to momentum conservation. This is why the
corrections derived in the previous Section diverge when $f$ goes to
unity.

When the weight keeps a constant sign, the correction due to momentum
conservation is of the same relative magnitude within the flow vector
method [Eq.~(\ref{flowQ+mom})] as within the two-particle method
[Eq.~(\ref{twoflow+mom})].  In the latter method, the correlation due
to flow is of order $v_1^2$ while the correlation due to momentum
conservation is of order $1/N$.  The flow vector method enhances both
terms by a factor of order $\sqrt{M}$, where $M$ is the number of
particles within the detector acceptance, which are used in
constructing the flow vector ${\bf Q}$.  Roughly speaking, this is
because the flow vector is analogous to a random walk of $M$ steps:
the length of the walk is approximately $\sqrt{M}$ times the size of a
step, and similarly the correlation between ${\bf Q}$ and a given
particle is $\sqrt{M}$ times the correlation between two particles.
This can be seen more explicitly from the formulas derived above.  We
first estimate the correlation due to flow, i.e., the first term in
the right-hand side of Eq.~(\ref{flowQ+mom}).  Since
$\mean{\cos\Delta\Phi}$ is proportional to $\chi$ for small $\chi$
(see Eq.~(\ref{resolution}) and the figures in
Refs.~\cite{Poskanzer:1998yz,Ollitrault:1997di}), and $\chi\sim
v_1\sqrt{M}$, this term is indeed of order $v_1^2\sqrt{M}$.  Let us
now estimate the second term, whose order of magnitude is given by
Eq.~(\ref{ordermag}).  We choose for simplicity a weight $w=p_T$
within the detector acceptance, and zero outside.  The parameter $f$
in Eq.~(\ref{deff}) then reduces to $f\simeq\sqrt{M/N}$.  Introducing
this value in the estimate Eq.\ (\ref{ordermag}), we find a
correlation of order $\sqrt{M}/N$, as expected.

The correction to the reaction plane resolution, on the other hand, is
generally of order $f^2$, which is roughly the fraction of the total
number of particles used in estimating the reaction plane.  
Note that the correction to the
resolution is not the same for all harmonics of the azimuthal
distribution.  Indeed, the resolution $\mean{\cos n\Delta\Phi}$ is
more sensitive to $\chi$ when $n > 1$, so that effects are larger for
higher harmonics relative to the first harmonic event plane.

Let us now discuss how the corrections vary with the centrality of the
collision.  The contribution of momentum conservation to the
correlation between a particle and the flow vector scales with the
number of particles $N$ like $1/N$.  Therefore, momentum conservation
may in fact dominate the measured correlation for the most peripheral
collisions.  For central collisions, the flow vanishes by symmetry and
only momentum conservation remains, but it is likely to be smaller due
to the larger value of $N$.  In contrast, the parameter $f$, which
determines the correction to the event plane resolution, is to a good
approximation independent of centrality. This is because the fraction
of particles in the detector acceptance varies little with centrality.
One could use this scaling property to estimate the contribution
of momentum conservation (and other nonflow effects)  experimentally.
This property was 
used
in Ref.~\cite{yidai}, where the contribution of momentum conservation
was estimated from the correlations between four subevents.

Besides analysis of the centrality dependence, one
can observe the contribution due to momentum conservation
by checking if $v_1(p_t,y)=0$ at midrapidity for all transverse
momenta.
In principle, the correlation of particles at midrapidity
with other
particles can be used to estimate the
momentum conservation effect. Checking the dependence on $p_T$ shown
in Eq.~(\ref{twomomentum}), could also be useful. 

It is important to notice that there still exist other
nonflow effects \cite{Poskanzer:1998yz}, as for instance correlations
due to resonance decays or quantum effects.  One can try to avoid them
by performing appropriate cuts in phase space or using carefully
chosen subevents \cite{Ackermann:2000tr}.  Alternatively, one can take
them into account by performing a similar correction to what we
propose here, through a detailed modeling of nonflow
effects~\cite{Borghini:2000cm,Dinh:1999mn}.  Finally, if statistics
are large enough, one can construct correlations between more than two
particles: then, nonflow effects can be eliminated in a model
independent way \cite{Borghini:2001vi}.

\section{Application to NA49 data}
\label{s:experiment}

To experimentally implement the correction we derived in the previous
sections, one should estimate the various coefficients which enter
Eq.\ (\ref{correl}).  Since all emitted particles are not detected,
one needs a parameterization of the cross section for each type of
emitted particle, and for each centrality bin, in order to compute the
average $\mean{p_T^2}$ for all particles and the total multiplicity
$N$.  One can then check that the results make sense by plotting the
rapidity dependence of the corrected directed flow, $v_1(y)$: if the
signal which remains after correction is indeed flow, it should cross
zero at midrapidity, since $v_1$ is expected to be an odd function of
$y-y_{CM}$.

\begin{figure}[hbt]
\caption{\label{fig:v1(y)} NA49 results on directed flow as a function of 
rapidity for charged pions from minimum bias 158$A$ GeV Pb + Pb.
Shown are $v_1$ before (squares) and after (circles) correction for
momentum conservation. Solid lines are polynomial fits.  The open
points have been reflected about midrapidity.}
\includegraphics*[width=\linewidth]{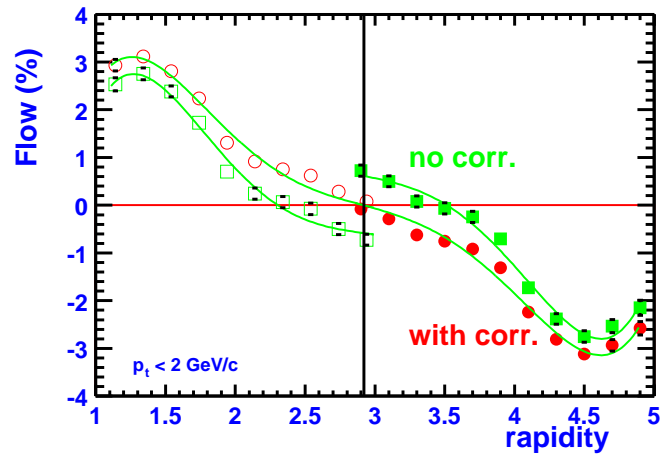}
\end{figure}

As an illustration, the method was applied to the NA49 data in Pb+Pb
collisions at 158 GeV per nucleon~\cite{Poskanzer:2001cx}.  In the
analysis, the event plane was determined from charged pions with $p_T
< 1.0$~GeV/$c$, except for centrality bins 1 and 2, where the upper
cuts were at 0.3 and 0.6~GeV/$c$, respectively. The $y$ values were
between 4.0 and 6.0. The particles were assigned weights $w=|y-y_{\rm
CM}|$, always positive since all used particles belong to the forward
hemisphere.  Then pions with known $p_T$ and $y$ values were
correlated with this event plane (being careful to remove
autocorrelations) to provide a matrix of observed $v_1(p_T,y)$.  The
$\chi_{\rm sub}$ values were obtained from the correlation of subevent planes
using Eqs.~(\ref{deff}) and (\ref{resolution2}).  In each $(p_T,y)$
bin the momentum conservation correction was subtracted using Eq.\
(\ref{correl}).  The full event plane resolution was obtained from
Eq.~(\ref{resolution}) and, according to Eq.\ (\ref{flowQ+mom}), the
whole matrix was corrected for the event plane resolution to produce
the double differential $v_1$ values.  The procedure was done for each
centrality bin separately and then the minimum bias results were
produced by taking the weighted average over centrality, weighting
with the cross section.  The cross sections as a function of $p_T$,
$y$, and centrality, had already been evaluated and parameterized
\cite{crosssections}.  Figure~\ref{fig:v1(y)} was produced by
projecting this final matrix onto the rapidity axis, by taking the
weighted average over $p_T$, again weighting with the cross section.

\begin{table}[hbt]
\caption {Listed for the six centralities are: $N$, the total
multiplicity over all phase space; $\mean{p_T^2}$, the mean $p_t^2$
over all phase space in units of GeV$^2/c^2$; $M$, the multiplicity of
particles used for the event plane; $f$, the fraction defined by
Eq.~(\ref{deff}); $\chi$, the resolution parameter for the full event
plane; the resolution of the full event plane; and the percent
increase in the resolution due to momentum conservation. The first two
centralities had more restrictive $p_T$ cuts.}
\label{tbl:NA49Table}
\begin{center}
\begin{ruledtabular}
\begin{tabular}{l|c|c|c|c|c|c} 
Centrality              & 1     & 2     & 3     & 4     & 5     & 6     \\ \hline
$N$                     & 2402  & 1971  & 1471  & 1028  & 717   & 457   \\
$\mean{p_T^2}$, GeV$^2/c^2$  
                        & 0.32  & 0.32  & 0.31  & 0.30  & 0.28  & 0.27  \\
$M$                     & 135   & 184   & 152   & 108   & 75    & 46    \\
$f$                     & 0.07  & 0.14  & 0.17  & 0.17  & 0.17  & 0.17  \\
$\chi$                  & 0.20  & 0.25  & 0.39  & 0.45  & 0.48  & 0.47  \\
resolution              & 0.18  & 0.22  & 0.33  & 0.38  & 0.40  & 0.39  \\
res. increase, \%       & 6.3  & 19.7 & 10.9 & 8.0 & 7.1 & 7.3
\end{tabular}
\end{ruledtabular}
\end{center}
\end{table}

The parameterized cross sections were also used to derive the
$\mean{p_T^2}$ for all particles in Eqs.\ (\ref{deff}) and (\ref{correl}).
More specifically, in each centrality bin the average was obtained by
taking 3/2 the sum of $p_T^2$ of the charged pions, plus twice the sum
for the charged kaons, plus the fraction of nucleons which are thought
to be participants, and then dividing by the total multiplicity.  The
values of the parameters used in this momentum conservation correction
are shown in Table~\ref{tbl:NA49Table}.  The $\mean{\sum p_T^2} = N
\mean{p_T^2}$ values range from 773 to 122 GeV$^2/c^2$; a
slightly higher value was used in Ref.~\cite{Borghini:2000cm},
resulting in a somewhat smaller correction. The last line of the table 
shows that the increase in resolution due to momentum conservation is
always less than 20\%.

However, Fig.\ \ref{fig:v1(y)} shows that the correction for the
effect of momentum conservation on $v_1$ is indeed significant,
roughly an absolute value of 1\% in the whole rapidity range under
study.  
The corrected correlation does
give a directed flow vanishing at midrapidity within error bars, as it
should. In fact, if one did not know $N$ and $\mean{p_T^2}$, one could
force the data to go through zero at midrapidity as a way to evaluate
this correction.

\begin{acknowledgments}
We would like to thank the NA49 Collaboration for permission to use
their data and, in particular, Alexander Wetzler for performing the
data analysis and Glen Cooper for parameterizing his cross section values.  
We also thank the anonymous referee for suggesting us the comparison with 
the method of Ref.\ \cite{ogilvie}. 
N.\ B.\ acknowledges support from the ``Actions deRecherche Concert{\'e}es'' 
of ``Communaut{\'e} Fran{\c c}aise de Belgique'' and IISN-Belgium.
\end{acknowledgments}

\vspace{0.5 cm} 
\appendix
\section{Correlation between one particle and the event flow vector}
\label{s:appA}

In this Appendix, we compute the azimuthal correlation between a
particle with momentum $p_T$ and the flow vector ${\bf Q}$ when both
flow and momentum conservation are taken into account.
Flow, which is defined as a correlation of outgoing momenta with the
reaction plane, means here a correlation with the $x$-axis: the
transverse momentum vector ${\bf p_T}$ in a given $y$ and $p_T$ window
has a non-zero average value $\mean{{\bf p_T}}$ which is parallel to
the $x$-axis, and the same holds for the flow vector ${\bf Q}$.  If
the only correlation was due to flow, we could write the normalized
$N$-particle momentum distribution as a product of single-particle
distributions:
\begin{equation}
\frac{dP}{d{\bf p_T}_1\cdots d{\bf p_T}_N}=
F_1({\bf p_T}_1)\cdots F_N({\bf p_T}_N),
\end{equation}
where $N$ denotes the total number of outgoing particles.  The
$F_j({\bf p_T}_j)$ are normalized to unity, and we further assume that
they satisfy average momentum conservation $\mean{{\bf
p_T}_1+\cdots+{\bf p_T}_N}=0$.

Here, we also take into account exact transverse momentum
conservation, which is expressed by means of an additional Dirac
constraint~\cite{Borghini:2000cm}:
\begin{widetext}
\begin{equation}
\label{pNcorr}
\frac{dP}{d{\bf p_T}_1\cdots d{\bf p_T}_N}=
\pi N\mean{p_T^2} F_1({\bf p_T}_1)\cdots F_N({\bf p_T}_N)
\delta\left({\bf p_T}_1+\cdots+{\bf p_T}_N\right).
\end{equation}
\end{widetext}
Note that $F_j({\bf p_T}_j)$ no longer strictly represents the
single-particle momentum distribution.  The latter is now given
by~\cite{Borghini:2000cm}:
\begin{equation}
\label{p1corr}
\frac{dP}{d{\bf p_T}_j}=F_j({\bf p_T}_j)\left(1-
\frac{p_{Tj}^2-\mean{p_{Tj}^2}}{N\mean{p_T^2}}\right).
\end{equation}

We are going to compute the correlated probability distribution
between the transverse momentum of an arbitrary particle (which we
choose to be the one labeled $N$, ${\bf p_T}_N$) and the flow vector
${\bf Q}$.  To compute this probability distribution $dP/d{\bf
p_T}_Nd{\bf Q}$, we introduce the flow vector ${\bf Q}$ defined by
Eq.~(\ref{Qvector}) in Eq.~(\ref{pNcorr}) by means of a Dirac
constraint, and integrate over the $N-1$ remaining particle momenta:
\begin{widetext}
\begin{equation}
\frac{dP}{d{\bf p_T}_N d{\bf Q}}=
\pi N\mean{p_T^2} F_N({\bf p_T}_N)\int\!
\left(\prod_{j=1}^{N-1}F_j({\bf p_T}_j)d{\bf p_T}_j\right)
\delta\left({\bf p_T}_1+\cdots+{\bf p_T}_N\right)
\delta\left({\bf w}_1+\cdots+{\bf w}_N-{\bf Q}\right),
\end{equation}
with ${\bf w}_j=w_j {\bf u}_j=(w_j/{p_T}_j){\bf p_T}_j$.

Using the Fourier representation of the Dirac distributions, this
equation becomes
\begin{equation}
\label{kl}
\frac{dP}{d{\bf p_T}_Nd{\bf Q}}=
\pi N\mean{p_T^2} F_N({\bf p_T}_N) \int\!\frac{d^2{\bf k}}{(2\pi)^2}
e^{i{\bf k}\cdot {\bf p_T}_N}
\int\frac{d^2{\bf l}}{(2\pi)^2}e^{-i{\bf l}\cdot{\bf Q}}
\prod_{j=1}^{N-1} \left(\int F_j({\bf p_T}_j)
e^{i({\bf k}\cdot {\bf p_T}_j+{\bf l}\cdot{\bf w}_j)}
d{\bf p_T}_j \right).
\end{equation}
We have assumed that the particle with momentum ${\bf p_T}_N$ is not
involved in the determination of the reaction plane, i.e., that ${\bf
w}_N=0$, as one usually does to avoid
autocorrelations~\cite{Danielewicz:1985hn}.

Since $N$ is large, the integrals over ${\bf k}$ and ${\bf l}$ can be
evaluated by means of a saddle-point approximation, i.e., only small
values of ${\bf k}$ and ${\bf l}$ contribute.  Expanding to second
order in ${\bf k}$ and ${\bf l}$ and re-exponentiating, one obtains
\begin{equation}
\label{saddlepoint}
\int F({\bf p_T})
e^{i({\bf k}\cdot {\bf p_T}+{\bf l}\cdot{\bf w})}d{\bf p_T}\simeq 
\exp\left(i{\bf k}\cdot\mean{\bf p_T}+i{\bf l}\cdot\mean{\bf w}
-\frac{{\bf k}^2}{4}\mean{p_T^2}
-\frac{{\bf l}^2}{4}\mean{w^2}
-\frac{{\bf k}\cdot{\bf l}}{2}\mean{wp_T}\right),
\end{equation}
\end{widetext}
where we have dropped the index $j$ for simplicity.  In deriving this
equation, we have assumed that directed flow is small for most
particles, $|v_1|\ll 1$, so that terms of second order in the flow can
be neglected, i.e. $\mean{\bf p_T}^2\ll \mean{p_T^2}$, etc.  Note that
the correlation between ${\bf p_T}_N$ and ${\bf Q}$ comes from the
cross term, proportional to ${\bf k}\cdot{\bf l}$.

Inserting Eq.~(\ref{saddlepoint}) into Eq.~(\ref{kl}), the remaining
integrals over ${\bf l}$ and ${\bf k}$ are Gaussian, and can thus be
evaluated analytically.  After some algebra, and dropping the index
$N$ for sake of brevity, one finally obtains the result in the form
\begin{eqnarray}
\label{corrdistribution}
&&\displaystyle \frac{dP}{d{\bf p_T} d{\bf Q}}= 
\displaystyle \frac{dP}{d{\bf p_T}} \frac{dP}{d{\bf Q}} \cr
&&\times\left(1+ 2\,c\left(\frac{{\bf p_T}-\mean{\bf p_T}}{p_T}\right)
\cdot \left(\frac{{\bf Q}-\mean{\bf Q}}{\sigma}\right)\right).
\end{eqnarray}
In this equation, $dP/d{\bf p_T}$ denotes the single-particle momentum
distribution [obtained by integrating Eq.~(\ref{corrdistribution})
over ${\bf Q}$], given by Eq.~(\ref{p1corr}), and $dP/d{\bf Q}$ the
distribution of the flow vector ${\bf Q}$ [obtained by integrating
Eq.~(\ref{corrdistribution}) over ${\bf p_T}$], given by Eq.\
(\ref{distri-Q}).  The dimensionless correlation strength $c$ is given
by
\begin{equation}
\label{defc}
c\equiv -\frac{p_T}{\sqrt{N\mean{p_T^2}}}\,\frac{f}{\sqrt{1-f^2}}, 
\end{equation}
where $f$ is given by Eq.~(\ref{deff}).  Finally, the width of the
distribution of ${\bf Q}$ is
\begin{equation}
\label{defsigma}
\sigma^2\equiv N\,\mean{w^2}\left(1-f^2\right).
\end{equation}
In this expression, $(1-f^2)$ is the correction to the width due to
momentum conservation.

The above expressions of $c$ and $\sigma$ have been obtained by a
``brute force'' calculation, but one can easily recover them directly
from the distribution (\ref{corrdistribution}), and the correlation
due to momentum conservation, Eq.~(\ref{twomomentum}), which we
rewrite in the form
\begin{equation}
\label{twomomentumbis}
\mean{{\bf u_1}\cdot{\bf u_2}}-\mean{{\bf u_1}}\cdot\mean{{\bf u_2}}
=-\frac{{p_T}_1{p_T}_2}{N\mean{p_T^2}}. 
\end{equation}
From the distribution of ${\bf Q}$, Eq.~(\ref{distri-Q}), one obtains
$\sigma^2=\mean{{\bf Q}^2}-\mean{\bf Q}^2$.  Using the definition of
${\bf Q}$, Eq.~(\ref{Qvector}), and Eq.~(\ref{twomomentumbis}), it is
a simple exercise to check Eq.~(\ref{defsigma}).  Similarly,
Eq.~(\ref{corrdistribution}) allows to express the correlation
strength $c$ in terms of simple averages: $c=(\mean{{\bf p_T}\cdot{\bf
Q}}-\mean{\bf p_T}\cdot\mean{\bf Q})/(\sigma p_T)$, which can also be
easily calculated using Eqs.~(\ref{Qvector}) and
(\ref{twomomentumbis}).

As explained in Sec.\ref{s:correction}, there are two ways of 
eliminating the correlation in Eq.~(\ref{corrdistribution}). 
The first possibility is to shift the flow vector for each 
particle. Making the change of variables 
${\bf Q'}={\bf Q}-c\sigma{\bf p_T}/p_T$ in the distribution 
Eqs.~(\ref{distri-Q}) and expanding to first order in the 
correlation $c$, one easily checks that the correlated term 
in Eq.~(\ref{corrdistribution}) cancels, so that the 
distributions of ${\bf Q'}$ and ${\bf p_T}$ are uncorrelated. 
Replacing $c$ and $\sigma$ by their expressions (\ref{defsigma}) 
and (\ref{defc}), one recovers Eq.~(\ref{shiftedQ}). 

Instead of redefining the flow vector, one can compute the azimuthal 
correlation between the particle and the flow vector 
using the correlated distribution 
Eq.\ (\ref{corrdistribution}). The first term in the right-hand side gives
the correlation due to flow, while the second term corresponds to the
correlation due to momentum conservation, that is, the term we wish to
calculate.  This term gives the following contribution:
\begin{equation}
\label{correl1}
\mean{\cos(\phi-\Phi)}_{\rm mom.\ cons.} = \frac{c}{\sigma}
\left(\mean{Q}-\mean{\cos\Delta\Phi}{\bf e}_x \cdot\mean{\bf Q}\right),
\end{equation}
The resolution $\mean{\cos\Delta\Phi}$ is given by
Eq.~(\ref{resolution}), and $\mean{Q}$ is obtained by a similar
calculation from the Gaussian distribution (\ref{distri-Q}):
\begin{equation}
\label{meanQ}
\mean{Q}=\sigma\,\frac{\sqrt{\pi}}{2}\,e^{-\chi^2/2}
\left[(1+\chi^2)I_0(\chi^2/2)+\chi^2 I_1(\chi^2/2)\right].
\end{equation}
The terms proportional to $I_1(\chi^2/2)$ cancel, and one finally
obtains
\begin{equation}
\label{correlpT-Q}
\mean{\cos(\phi-\Phi)}_{\rm mom.\ cons.} =
c\, \frac{\sqrt{\pi}}{2}e^{-\chi^2/2}I_0(\chi^2/2). 
\end{equation}
In the limit $\chi\ll 1$ (poor resolution), this equals approximately
$c\sqrt{\pi}/2$, while for $\chi\gg 1$ (good resolution), the
correction decreases like $c/(2\chi)$.

\section{Correlation between two subevents}
\label{s:appB}

We shall now compute the azimuthal correlation between two subevents,
which we denote ${\bf Q_a}$ and ${\bf Q_b}$, in the presence of flow
and momentum conservation.  For simplicity, we assume throughout the
following discussion that both subevents are equivalent, and moreover
that they are of equal multiplicities.  Since both contain a large
number of particles, we may directly apply the central limit theorem.
The normalized correlated distribution is thus Gaussian:
\begin{widetext} 
\begin{equation}
\label{distsubev}
\frac{dP}{d{\bf Q_a}d{\bf Q_b}} = \frac{1}{\pi^2\sigma_a^4(1-C^2)} \exp \left(
-\frac{({\bf Q_a}-\mean{\bf Q_a})^2 + ({\bf Q_b}-\mean{\bf Q_b})^2 - 
2C({\bf Q_a}-\mean{\bf Q_a}) \cdot ({\bf Q_b}-\mean{\bf Q_b})}{\sigma_a^2 
(1-C^2)} \right). 
\end{equation}
\end{widetext}
In this equation, $C$ represents the strength of the ``nonflow''
correlation (in particular, that due to momentum conservation) between
the subevents.  For $C=0$, subevents are independent, and the
distribution (\ref{distsubev}) factorizes into the product of two
distributions of the type (\ref{distsubeva}).  In the more general
case $C\not= 0$, one recovers Eq.~(\ref{distsubeva}) by integrating
Eq.\ (\ref{distsubev}) over ${\bf Q_b}$.

When correlations due to momentum conservation are taken into account,
the width $\sigma_{\rm sub}^2 =\mean{{\bf Q_{\rm sub}}^2}-\mean{\bf Q_{\rm sub}}^2$ is given
by a formula analogous to Eq.\ (\ref{defsigma}), with the important
difference that $N$ and $f^2$ are replaced by $N/2$ and $f^2/2$, since
we use only half the particles to construct the subevents:
\begin{equation}
\label{defsigmaa}
\sigma_{\rm sub}^2\equiv \frac{N}{2}\,\mean{w^2}\left(1-\frac{f^2}{2}\right).
\end{equation}
Note that $\sigma_{\rm sub}$ is not simply equal to $\sigma/\sqrt{2}$, because
of the term arising from correlations.

Let us now calculate the correlation strength $C$.  It can be obtained
by taking the following average value with the probability
distribution (\ref{distsubev})~\cite{Poskanzer:1998yz,Ollitrault:dy}:
\begin{eqnarray}
\label{defC}
C & = & \frac{
\mean{({\bf Q_a}-\mean{\bf Q_a}) \cdot ({\bf Q_b}-\mean{\bf Q_b})}}{
\sigma_a^2}\cr
 & = & \frac{\mean{{\bf Q_a}\cdot{\bf Q_b}} - 
\mean{\bf Q_a}\cdot\mean{\bf Q_b}}{\sigma_a^2}.
\end{eqnarray}

Using the definitions of ${\bf Q_a}$ and ${\bf Q_b}$ and the
correlation from momentum conservation, Eq.~(\ref{twomomentumbis}),
the numerator of Eq.\ (\ref{defC}) is equal to $-(N/2)^2
\mean{wp_T}^2/N\mean{p_T^2}$, where we have neglected terms of second
order in flow.  With the width given by Eq.\ (\ref{defsigmaa}), we
finally obtain
\begin{equation}
\label{valC}
C = \frac{-\mean{wp_T}^2}{2\mean{p_T^2}\mean{w^2}(1-f^2/2)} = 
\frac{-f^2}{2-f^2}, 
\end{equation}
where $f$ is given by Eq.\ (\ref{deff}). 

As explained in Sec.\ \ref{s:standard}, the resolution parameter
$\chi_{\rm sub}=\langle {\bf Q_{\rm sub}}\rangle\cdot {\bf e_x}/\sigma_{\rm sub}$ of the
subevent is obtained from the distribution of the relative angle
between ${\bf Q_a}$ and ${\bf Q_b}$.  Unfortunately, there is no
general analytic expression for this distribution.

Useful expressions can be obtained in a few limiting cases.  First, if
the correlation strength $C$ is much smaller than unity, the
distribution (\ref{distsubev}) can be expanded to first order in $C$,
which yields a distribution analogous to Eq.~(\ref{corrdistribution}).
From this expression, one can in particular evaluate the average
cosine $\mean{\cos(\Phi_a-\Phi_b)}$, which leads to
Eq.~(\ref{resolution2}).  Second, if the resolution is high
($\chi_{\rm sub}\gg 1$), both ${\bf Q_a}$ and ${\bf Q_b}$ are close to their
average value $\mean{\bf Q_{\rm sub}}$, so that their azimuthal angles with
respect to the reaction plane are small, and one may approximate
$\Phi_a\simeq {\bf Q_a}\cdot {\bf e_y}/|\mean{\bf Q_a}|$, and a
similar expression for $\Phi_b$. In this limiting case, the
distribution of $\Phi_a$ and $\Phi_b$ is a Gaussian, and the
distribution of the relative angle is given by
Eq.~(\ref{highresolution}).

Quite remarkably, one can derive an exact expression, valid for
arbitrary $\chi_{\rm sub}$ and $C$, of the fraction of events for which the
angle between subevents is larger than $\pi/2$, taking into account
both flow and momentum conservation.  The condition
$|\Phi_a-\Phi_b|>\pi/2$ can equivalently be written ${\bf
Q_a}\cdot{\bf Q_b}<0$, so that
\begin{equation}
P(|\Phi_a-\Phi_b|>\pi/2) = \int{d{\bf Q_a} d{\bf Q_b}
\frac{dP}{d{\bf Q_a} d{\bf Q_b}}\,\theta(-{\bf Q_a}\cdot{\bf Q_b})}.
\end{equation}
We then insert the Fourier representation of Heaviside's step function
$\theta(x)$:
\begin{equation}
\label{heaviside}
\theta(-x)=\frac{i}{2\pi}
\lim_{\varepsilon\to 0^+}
\int_{-\infty}^{+\infty}\frac{e^{2 i k x}}{k+i\varepsilon}dk.
\end{equation}
With the probability distribution Eq.~(\ref{distsubev}), the integrals
on ${\bf Q_a}$ and ${\bf Q_b}$ are Gaussian and can be evaluated
analytically. One obtains
\begin{widetext}
\begin{equation}
\label{dir}
P(|\Phi_a-\Phi_b|>\pi/2) = \frac{i}{2\pi}\int_{-\infty}^{+\infty}
\frac{dk}{(k+i\varepsilon)\left[1-ik(1+C)\right]\left[1+ik(1-C)\right]}
\exp\left[\frac{2i\chi^2 k}{1-ik(1+C)}\right]
\end{equation}
\end{widetext}
The only essential singularity of the integrand occurs in the lower
half plane at $k=-i/(1+C)$.  Therefore the integral can be calculated
by closing the contour on the upper half plane. The only pole is at
$k=i/(1-C)$, and the theorem of residues yields~\cite{Ollitrault:dy}
\begin{equation}
\label{Pacute}
P(|\Phi_a-\Phi_b|>\pi/2)=\frac{1-C}{2}e^{-\chi_{\rm sub}^2}.
\end{equation}
Inserting the correlation strength Eq.~(\ref{valC}) in this
expression, one recovers Eq.~(\ref{obtuscorr}).

\end{document}